\documentstyle[11pt]{article}
\begin{document}
\setlength{\oddsidemargin}{0cm}
\setlength{\evensidemargin}{0cm}
\baselineskip=20pt

\hfill NIM-2001-08

\begin{center} {\Large\bf The Happer's Puzzle Degeneracies and Yangian}\end{center}

\bigskip

\begin{center}  { \large Cheng-Ming ${\rm Bai}^{1}$},\ \  {Mo-Lin ${\rm Ge}^{1}$ and Kang ${\rm Xue}^{1, 2}$}\end{center}

\begin{center}{\it 1. Liuhui Center for Applied Mathematics and Theoretical Physics Division, Nankai Institute of Mathematics, Nankai University, Tianjin 300071, P.R. China}\end{center}

\begin{center}{\it 2. Dept. of Physics, Northeast Teaching University, Chang Chun, Jilin, 130024, P.R. China}\end{center}

\begin{center} {\large\bf   Abstract } \end{center}

We find operators distinguishing the degenerate states for the Hamiltonian 
$H= x(K+\frac{1}{2})S_z +{\bf K}\cdot {\bf S}$ at $x=\pm 1$ that was given by
Happer et al$^{[1,2]}$ to interpret the curious degeneracies of the Zeeman effect for condensed vapor of $^{87}$Rb. The operators obey Yangian commutation relations. We show that the curious degeneracies seem to verify the Yangian algebraic structure for quantum tensor space and are consistent with the representation theory of $Y(sl(2))$.

PACS numbers: 02.20,03.65,33.35+r 

\newpage

{\bf(I) Introduction}\quad There occur curious degeneracies observed in the experiment for condensed  vapor of $^{87}$Rb and $^{85}$Rb$^{[1]}$ that are converted into ``anti-level-crossing'' due to a spin-axis interaction for triplet ($S=1$). To describe the Hamiltonian of a triplet dimer neglecting the quadrapole interaction, Happer et al introduced $^{[1,2]}$
$$H={\bf K}\cdot {\bf S}+x(K+\frac{1}{2})S_z, \eqno (1)$$
and pointed out that when $x=\pm 1$ there appear the puzzle degeneracies for $S=1$, where ${\bf K}$ and ${\bf S}$ are angular momentum and spin, respectively, ${\bf K}^2=K(K+1)$ and ${\bf S}^2=S(S+1)$. In Ref. [2] the eigenvectors corresponding to $E=-\frac{1}{2}$ had been given and a lot of elegant discussions were made. However, what is behind the curious degeneracy and why it happens only for $S=1$$^{[2]}$ are still missing.

Note that the Hamiltonian Eq.(1) works in a quantum tensor space $\Omega=V_{S}\otimes V_{K}$. If the $S_z$ term is replaced by $G_z$ in Eq.(1), through the C-G coefficients $\Omega$ can be decomposed into block-diagonal form $\Omega=\sum \Omega_{G}$, where ${\bf G}={\bf K}+{\bf S}$, i.e. instead of the whole tensor space $\Omega$, if each block is studied, then $\Omega$ has been well studied. This is the Lie-algebraic description. The degenerate states with different $G_z=m$ can be designated by acting $G_+$ on the given lowest ($m=-G$) state, or by $G_-$ on highest ($m=G$) one, where ${\bf G}^2=G(G+1)$. Hence $G_+$ ($G_-$) plays raising (lowering) role to designate the degenerate states.

Now for the Hamiltonian Eq.(1) the tensor space spanned by ${\bf S}$ and ${\bf K}$ is no longer block diagonal, i.e. the space $\Omega$ serves as a whole entity. Actually, the eigenvector of $H$ with any allowed $m$ (except $m=K+1$ and $m=-K$) in the degeneracy set $\alpha_{Dm}$ given in Ref. [2]:
$$H\alpha_{Dm}=-\frac{1}{2}\alpha_{Dm}\;\;(E_{Dm}=-\frac{1}{2}),\eqno (2)$$
is $^{[2]}$ 
$$ \alpha_{Dm}=-\sqrt{ \frac{(K+1+m)(K+1-m)}{4(K+\frac{1}{2})(K+m+\frac{1}{2})}}\alpha_1
+\sqrt{\frac{(K+1+m)(K+m)}{2(K+\frac{1}{2})(K+m+\frac{1}{2})}}\alpha_2
+\sqrt{\frac{(K+m)(K-m)}{4(K+\frac{1}{2})(K+m+\frac{1}{2})}}\alpha_3, \eqno (3)$$
where for given conserved $m=K_z+S_z$, $\alpha_1$, $\alpha_2$ and $\alpha_3$ are eigenvectors with $S_z=1,\;0,\;-1$ and $K_z=m-1,\;m,\;m+1$, respectively. Obviously $\alpha_{Dm}$ are no longer the eigenstates of Lie algebra $sl(2)$. Note that in  Eq.(3) the $\alpha_{Dm}$ occupies the whole indecomposable tensor space and possesses $(2K+1)$ generated states.  Are there still any raising and lowering operators that distinguish $\alpha_{Dm}$, like $G_{\pm}$ do in the usual degeneracies? In this paper, we shall give such operators and show their properties. Surprisingly we find such operators had been ready in mathematical physics, called Yangian algebra which is a simple extension of Lie algebra, and firstly presented by Drinfeld$^{[3]}$ in dealing with Yang-Baxter system $^{[4-5]}$. The representation theory of Yangian was also established $^{[6]}$. Actually Yangian had been employed in Haldane-Shastry Model $^{[7]}$.

In this paper we shall show how the Yangian algebra demonstrates the puzzle degeneracies and why they occur only for $S=1$. 

For the late convenience we use the new notations: $^{[6]}$
$$ \alpha_{Dp}={[2(2K+1)(p+\frac{1}{2})]}^{-\frac{1}{2}}
[-\sqrt{(p+1)(2K+1-p)}\alpha_1 
+\sqrt{2p(p+1)}\alpha_2+\sqrt{p(2K-p)}\alpha_3]\eqno (4)$$
where
$$\alpha_1=e_2\otimes e_{p-1},\;\alpha_2=\sqrt{\frac{p}{2(2K-p+1)}}e_1\otimes e_p,\;\alpha_3=\sqrt{\frac{p(p+1)}{(2K-p+1)(2K-p)}}e_0\otimes e_{p+1},\eqno (5)$$
$$e_2\otimes e_{p-1}=|1,p-1-K\},\;e_1\otimes e_p=|0,p-K\},\;e_0\otimes e_{p+1}=|-1, p+1-K\},\eqno (6)$$
$$p=m+K,\;\;p=-1,\;0,\;1,\;2,\;\cdots,\;2K,\;2K+1.\;\eqno (7)$$
The action on the new basis is given by$^{[6]}$
\begin{eqnarray*}
&&x^+e_{2K}=0, x^+e_p=(p+1)e_{p+1},0\leq p<2K;x^-e_0=0,x^-e_p=(2K-p+1)e_{p-1},0<p\leq 2K;\\
&&h e_p=(p-K)e_p,0\leq p\leq 2K,\hspace{10.2cm} (8)
\end{eqnarray*}
here $x^{\pm}=S_{\pm}$ or $K_{\pm}$ and $h=S_z$ or $K_z$ depending on the first space where spin operator acts on or second one where ${\bf K}$ acts on and  $e_0,\;e_1,\;e_2$ are the spin eigenvectors with $S_z=-1,\;0,1$, respectively. 
The Hamiltonian Eq.(1) for $x=1$  still has the property: $H\alpha_{Dp}= E_{Dp} \alpha_{Dp}, E_{Dp}=-\frac{1}{2}$.

{\bf (II) Raising operation for $\alpha_{Dm}$}\quad The eigenvectors of Hamiltonian Eq.(1) have three sets $\alpha_{Dm},\alpha_{Tm},\alpha_{Bm}$ as given by Happer$^{[2]}$ that form the whole Hilbert space, but the degeneracies occur in the set $\alpha_{Dm}$ only. Let us focus on this set possessing $E_{Dm}=-\frac{1}{2}$. The result in Ref. [2] can be briefly  illustrated by Table 1 where the linear combination for the states with $G=K+1,K,K-1$ is understood.

{\small $$\matrix{  & G=K+1 & G=K & G=K-1 &             & D-{\rm set} & T-{\rm set} & B-{\rm set} \cr
m=K+1 & ---  &     &       & \rightarrow &       & \alpha_{T,m=K+1} &\cr
m=K    & --- & ---   &       &\rightarrow & \alpha_{D,m=K}& \alpha_{T,m=K} &\cr
 m=K-1 & --- & --- & --- & \rightarrow  &\alpha_{D,m=K-1}& \alpha_{T,m=K-1} &\alpha_{B,m=K-1}\cr
 \vdots & \vdots &\vdots & \vdots &\vdots &\vdots &\vdots &\vdots\cr
m &--- &--- &--- &\rightarrow & \alpha_{Dm} &\alpha_{Tm} &\alpha_{Bm}\cr
\vdots & \vdots &\vdots & \vdots &\vdots &\vdots &\vdots &\vdots\cr
m=-K+1 & --- & --- & --- & \rightarrow  &\alpha_{D,m=-K+1}& \alpha_{T,m=-K+1} &\alpha_{B,m=-K+1}\cr
m=-K    & --- & ---   &       &\rightarrow & & \alpha_{T,m=-K} &\alpha_{B,m=-K}\cr
m=-K-1 & ---   &     &       & \rightarrow &      \alpha_{D,m=-K-1} & &\cr
&&&&&&&\cr
&&&&{\rm Table 1}&&&\cr}$$}

There are in total $(2K+1)$ eigenvectors of $H$ with $E_{Dm}=-\frac{1}{2}$ (see Table 1). In $D$-set, there are no states with $m=K+1$ and $m=-K$($m=-K-1$ and $m=K$ for $x=-1$). In particular, there is a curious ``gap'' between $\alpha_{D,m=-K+1}$ and $\alpha_{D,m=-K-1}$. Besides $D$-set (with degeneracies) there are other two, i.e. $T$-set and $B$-set that are without degeneracy. If we imagine $D$, $T$ and $B$ as three ``directions'', the $D$-``direction'' represents the set of degenerated states.

In order to introduce the raising or lowering operator to 
designate the degeneracy states in $D$-set, we recast Eq.(5) to 
$$\alpha_{Dp}=\frac{p}{\sqrt{2(2K+1)(p+\frac{1}{2})}}\sqrt{\frac{p+1}{2K+1-p}}
[-(2K+1-p) e_2\otimes e_{p-1}+pe_1\otimes e_p+pe_0\otimes e_{p+1}].\eqno (9)$$
It is easy to see that the action of $G_+=K_++S_+$ does not preserve the $D$-set.
A natural extension is the linear combination of ${\bf K}$ and ${\bf S}$, say $\mu {\bf K}+\lambda{\bf S}$. However, by acting it on $\alpha_{Dp}$, the resultant state will go beyond $\alpha_{Dp}$, i.e. will reach $T$-set and $B$-set. Actually, if we define $T_+=\lambda S_+ +\mu K_+$, then the demand $T_+\alpha_{Dp}\sim \alpha_{Dp+1}$ leads to $\lambda=\mu$ and $K=-\frac{1}{2}$ or $\lambda=\mu=0$. However, the quantum number $K$ must be positive, hence $T_+$ cannot be the raising operator. 

Thus, we have to look for a new candidate. Let
$$J_+=aS_++bK_++\frac{1}{2}(S_zK_+ - S_+K_z),\eqno (10)$$
and next we shall prove that by the appropriate choice of $a$ and $b$ in Eq.(10) the operator $J_+$  is really the raising operator preserving $D$-set. 

Acting $J_+$ on $e_0\otimes e_p,\;e_1\otimes e_p$ and $e_2\otimes e_p$, we obtain
$J_+(e_0\otimes e_p)= (a-\frac{p}{2}+\frac{K}{2})e_1\otimes e_p+(b-\frac{1}{2})(p+1)e_0\otimes e_{p+1}$,
$J_+(e_1\otimes e_p)=(2a-p+K)e_2\otimes e_p+b(p+1)e_1\otimes e_{p+1}$,$ J_+(e_2\otimes e_p)= (b+\frac{1}{2})(p+1)e_2\otimes e_{p+1}$.
Putting all the above relations together we have
\begin{eqnarray*}
J_+\alpha_{Dp}&\sim &p(p+2)(b-\frac{1}{2})\{ e_0\otimes e_{p+2}+\frac{b(p+1)+a-\frac{p}{2}-\frac{1}{2}+\frac{K}{2}}{(b-\frac{1}{2})(p+2)}e_1\otimes e_{p+1}\\
&& +\frac{[-(b+\frac{1}{2})(2K+1-p)+2a-p+K]}{(b-\frac{1}{2})(p+2)}e_2\otimes e_{p}\}.\hspace{5.5cm} (11)
\end{eqnarray*}
In order to identify Eq.(11) with Eq.(9) for $p\rightarrow p+1$ regardless the renormalization constant we should have
$$\frac{b(p+1)+a-\frac{p}{2}-\frac{1}{2}+\frac{K}{2}}{(b-\frac{1}{2})(p+2)}=1\eqno (12)$$
$$\frac{(b-\frac{1}{2})(p+2)}{-(b+\frac{1}{2})(2K+1-p)+2a-p+K}=-\frac{p+1}{2K-p}\eqno (13)$$
Eq.(12) and Eq. (13) give
$$a-b=-\frac{K}{2}-\frac{1}{2}\eqno (14)$$
where
$$a=-\frac{K}{2}+p+1=m+\frac{K}{2}+1,\;b=p+\frac{3}{2}=m+K+\frac{3}{2},\;{\rm for}\;-(K-1)\leq m\leq K-1,\eqno (15)$$
i.e. with the choice of the $a$ and $b$ satisfying Eq.(15) we have

$J_+\alpha_{Dm}\longrightarrow \alpha_{Dm+1}$, for $-K+1\leq m\leq K-1$ \hfill (16)

\noindent that preserves the set $\alpha_{Dm}$ for $|m|\leq K-1$. It is interesting to note that with the given parameters $a$ and $b$ for $-K+1\leq m\leq K-1$ we have $J_+\alpha_{D,m=K-1}\rightarrow \alpha_{D,m=K}$, i.e. the raising operator $J_+$ automatically guarantees to reduce the combination of three states to that of two states. 

Further, acting the extrapolated $J_+$ with $m=K$ on $\alpha_{D,m=K}$ we have $J_+\alpha_{D,m=K}=0$. The state $\alpha_{D,m=-K}$ should take the form $\alpha_{D,m=-K}=ue_1\otimes e_0+ve_0\otimes e_1$. It is easy to verify that $H\alpha_{D,m=-K}=Kv e_1\otimes e_0+(-\frac{3}{2}v+u)e_0\otimes e_1$. Hence $H\alpha_{D,m=-K}=-\frac{1}{2}\alpha_{D,m=-K}$ if and only if $K=-\frac{1}{2}$ and $u=v$ or $u=v=0$, i.e. $\alpha_{D,m=-K}$ is not an eigenvector of $H$. However, the state  $\alpha_{D,m=-K-1}=e_0\otimes e_0$ is. We find for $a=-\frac{K}{2}$ and $b=\frac{1}{2}$ given by Eq.(15) for the extrapolated $m=-K-1$:
$$J_+\alpha_{D,m=-K-1}=(a+\frac{K}{2})e_1\otimes e_0+(b-\frac{1}{2})e_0\otimes e_1=0.\eqno (17)$$
Amazingly $J^+$ with the given $a$ and $b$ by Eq. (15) automatically works for all $m$ and excludes the state $\alpha_{D,m=-K}$, and preserves the set of eigenstates for $E_D=-\frac{1}{2}$. To reach $\alpha_{D,m=-K+1}$ from $\alpha_{D,m=-K-1}$, we can act  the composition operator $J_+G_+$ ($G_+\alpha_{D,m=-K-1} =e_1\otimes e_0+e_0\otimes e_1$) on $\alpha_{D,m=-K-1}$ with
$a=-\frac{K}{2}-\frac{1}{2}$ and $b=\frac{1}{2K}+\frac{1}{2}=\frac{1}{2}(1+K^{-1})$. That together with Eq.(15) form the $J_+$ sequence to designate  all the 
states in $D$-set.

{\bf (III)\quad Lowering operators}\quad The lowering operator is
$$J_-=cS_-+dK_--\frac{1}{2}(S_zK_--S_-K_z)\eqno (18)$$
To preserve the $D$-set we find the solution:
$$c-d=\frac{K}{2},\; c=-\frac{K}{2}+\frac{1}{2}-m,\;d=\frac{1}{2}-m-K,\;{\rm for}\;-K+1\leq m\leq K-1.\eqno (19)$$
It is interesting to note that by acting the extrapolated $J_-$ with $m=-K+1$ on $\alpha_{D,m=-K+1}$, we find $J_-\alpha_{D,m=-K+1}=0$ that automatically exclude the state $\alpha_{D,m=-K}$ from the set of eigenstates for $E_{D}=-\frac{1}{2}$.

As pointed out in the section (II), the raising operator $J_+$ automatically generates the state $\alpha_{D,m=K}$. Now $J_-$ automatically makes truncation at $\alpha_{D,m=-K+1}$. It is believed that these operators are not occasionally introduced to describe the puzzle degeneracies. Furthermore, acting the extrapolated $J_-$ on $\alpha_{D,m=K}$, we have $J_-\alpha_{D,m=K}\sim \alpha_{D,m=K-1}$ if and only if $c=\frac{1}{2}-\frac{3}{2}K,\; d=\frac{1}{2}-2K$ which is just given by Eq.(19) for $m=K$.

The introduced $J_-$ is independent of the conjugate of $J_+$. However there is relation for the parameters $a-b=-\frac{K}{2}-\frac{1}{2}$ and $c-d=\frac{K}{2}$ that has deep implication in the representation theory of Yangian. We shall discuss it later. The introduced $J_+$ and $J_-$ work perfectly to coincide with the curious degeneracies and look larger than angular momentum theory. There should be something behind the game.

{\bf (IV)\quad Yangian algebra}\quad Regardless the particular choice of parameters $a,b,c$ and $d$, let us consider the operator
$${\bf J}=a'{\bf S}+b'{\bf K}-\frac{\sqrt{-1}}{2}{\bf S}\times {\bf K},\eqno (20)$$
where $a'$ and $b'$  are free parameters. Obviously Eq.(20) gives $J_+,J_-$ appearing in Eq. (10) and Eq. (18) and $J_3$. The commutation relations for ${\bf J}$ and the total angular momentum ${\bf I}={\bf G}={\bf S}+{\bf K}$ form the so-called Yangian algebra associated with $sl(2)$.
The parameters $a'$ and $b'$ play important role in the representation theory of Yangian given by Chari and Pressley $^{[6]}$. The set $\{{\bf I},\;{\bf J}\}=Y(sl(2))$ obeys the commutation relations of $Y(sl(2))$ ($A_{\pm}=A_1\pm \sqrt{-1}A_2$):
$$[I_3,\;I_{\pm}]=\pm I_{\pm},\;\;[I_+,\;I_-]=2I_3,\;\;\;(sl(2));\eqno (21)$$
$$[I_3,\;J_{\pm}]=[J_3,\;I_{\pm}]=\pm J_{\pm},\;\;[I_+,\;J_-]=[J_+,\;I_-]=2J_3,\eqno (22)$$
(i.e. $[I_i,\;J_j]=\sqrt{-1}\varepsilon_{ijk}J_k$) and nonlinear relation
$$[J_3,\;[J_+,\;J_-]]=\frac{1}{4}I_3(I_+J_- -J_+I_-)\eqno (23)$$
that forms an infinitely dimensional algebra generated by 6 generators. All the other relations given in Ref.[3] can be obtained from Eq.(21)-Eq.(23) together with the Jacobian identities.$^{[8]}$

The essential difference between the representations of Yangian algebras and those of Lie algebras is the appearance of the free parameters $a'$ and $b'$ whose originally physical meaning is one-dimensional momentum. A special choice corresponds to a particular model. Applying the Yangian representation theory to Hydrogen atom, it yields the correct spectrum ($\sim n^{-2}$) that is the simplest example of the consistence between Yangian and Quantum Mechanics $^{[9]}$. Now the Happer's degeneracies can be viewed as another example. Furthermore, we would like to make the following remarks:

(a) The elements of $J_+$ given by Eq.(10) $< \alpha_{Dm'}|J_+|\alpha_{Dm}>\sim
<\alpha_{Dm'}|K_+|\alpha_{Dm}>\ne 0$, because $<\alpha_{Dm'}|{\bf S}|\alpha_{Dm}>=<\alpha_{Dm'}|{\bf S}\times {\bf K}|\alpha_{Dm}>=0$, as pointed out in Ref. [2] (see Eq.(2.23) in Ref. [2]). This indicates that the role played by $J_+$ in the ``$D$-direction'' likes the role played by $K_+$. Why do we need Yangian?  The terms of $S_+$ and $({\bf K}\times {\bf S})_+$ should be added to guarantee $<\alpha_{Tm'}|J_+|\alpha_{Dm}>=$$<\alpha_{Bm'}|J_+|\alpha_{Dm}>=0$, namely, if only acting $K_+$ on $\alpha_{Dm}$ it yields non-vanishing transitions to $\alpha_{Tm'}$ and $\alpha_{Bm'}$ that  no longer preserves the $D$-set. The part other than $K_+$ in Yangian $J_+$ given by Eq.(10) exactly cancel the nonvanishing contribution received from ``$T$-'' and ``$B$-direction''.

(b) Observing the process determining parameters $a$ and $b$ in Eq.(12) and Eq.(13), the reason for the existence of solution of $a$ and $b$ is clear. For $S=1$, the eigenvector of $ H$ is formed by three base. Regardless over-all normalization factor there are two independent coefficients. In requiring $J_+\alpha_{Dm}\sim \alpha_{Dm+1}$, we have to make comparison between the coefficients of the independent base in $J_+\alpha_{Dm}$ and $\alpha_{Dm+1}$ to determine the unknown parameters $a$ and $b$. For spin $S=1$, there are just two equations for $a$ and $b$. However, for spin $S>1$, in general, one is unable to find solution for $a$ and $b$ to fit the number of equations larger than two. Therefore the Yangian description of the curious degeneracies admits only $S=1$ for arbitrary $K$. This is consistent with experiment.$^{[1,2]}$

(c) In fact, the parameters appearing in $J_+$ and $J_-$ exactly coincide with the conditions of existence of the subrepresentations of Yangian$^{[6]}$. Following the theorem in [6], for $a-b=-\frac{K}{2}-\frac{1}{2}$, the subspace spanned by vectors with $G=K+1$ is the unique irreducible subrepresentation of $Y(sl(2))$, that is, the states with $G=K+1$ are stable under the action of ${\bf J}$. In particular, for the given $a$ and $b$ in Eq. (15), the action of $J_+$ on the states with $G=K+1$ is given by $J_+\alpha_{G=K+1,m}=(m+K+1)G_+\alpha_{G=K+1,m}$ and at the same time, $J_+$ will make the states with $G=K$ and $G=K-1$ transit to $G=K+1$, but not vice versa, called ``directional transition''$^{[8]}$, i.e. the transition given rise by Yangian goes in one way. For $c-d=\frac{K}{2}$, $G=K-1$ is the unique irreducible subrepresentation and for $c$ and $d$ given by Eq. (19), acting $J_-$ on the states with $G=K-1$, we have $J_-\alpha_{G=K-1,m}=-(m+K)G_-\alpha_{G=K-1,m}$. Therefore the representation theory of $Y(sl(2))$ tells that the relationship between $a-b$ and $c-d$ given by Eq. (14) and Eq. (19), respectively, should be held to preserve the states with $G=K+1$ (or $G=K-1$) possessing Lie algebraic behavior.

(d) We have seen that the $J_-$ is not the conjugate of $J_+$. Such a phenomenon is reasonable because $\alpha_{Dm}$ is neither the Lie-algebraic state nor symmetry of $H$. Hence, the $D$-set is not a subrepresentation, i.e., $D$-set cannot be stable under all the actions of ${\bf J}$, but stable under $J_+$ and $J_-$ with the different parameters which just satisfy the condition for subrepresentation of Yangian.

(e) The third component of ${\bf J}$ takes the form $J_3 =aS_z+bK_z+S_+K_- - S_-K_+ $. For any parameters, the action of $J_3$ will not keep the $D$-set. But, with the suitable $a-b=1$, the operator $J_3+2(2K+1)S_z^2$ will keep the $D$-set.

(f) The Eq. (1) looks like describing a ``dimer pair'' with spin $1$, relative angular momentum ${\bf L}=0$ and orbital angular momentum ${\bf K}$, where ${\bf S}+{\bf L}+{\bf K}$ is conserved. It is expected to introduce the Yangian to discuss an orbital rotated spin $1$-pair with relative ${\bf L}=0$, in contrast to the usual $p$-pair
with ${\bf K}=0$.$^{[10]}$

In conclusion we have read of a new type of algebra structure(Yangian) from the Happer's degeneracies and such an algebra had been ready by Drinfeld$^{[3]}$. All the analysis coincides with the representation theory of $Y(sl(2))$$^{[6]}$ for the special choice of $a,b$ in $J_+$ and $c,d$ in $J_-$. It also leads to the fact that only $S=1$ is allowed to yield the curious degeneracies. If Zeeman effect tells Lie algebra, then the curious degeneracies may tell the existence of Yangian. 

{\bf Acknowledgements}

We would like to thank Freeman Dyson who guided the authors to deal with this problem and W. Happer for encouragement and enlightening communications. This work is in part supported by NSF of China. 

\baselineskip=15pt
{\bf References}
\begin{description}
\item[[1]] C.J. Erickson, D. Levron, W. Happer, S. Kadlecek, B. Chann, L.W. Anderson, T.G. Walker, Phys. Rev. Lett. 85, 4237 (2000).
\item[[2]] W. Happer, Degeneracies of the Hamiltonian $x(K+1/2)S_z+{\bf K}\cdot {\bf S}$, preprint, Princeton Univ., Nov, 2000.
\item[[3]] V. Drinfeld, Sov. Math. Dokl 32, 254 (1985); 36, 212(1988);Quantum groups, in PICM, Berkley, 269 (1986). 
\item[[4]] L.D. Faddeev, Les Houches, Session 39, 1982.
\item[[5]] E.K. Sklyanin, Quantum Inverse Scattering Methods, Selected Topics, in M.L. Ge (ed.) Quantum Groups and Quantum Integrable Systems, World Scientific, Singapore, 63-88 (1991).
\item[[6]] V. Chari and  A. Pressley , A guide to quantum groups, Cambridge University Press, 1994; Yangian and $R$-matrix, L'Enseignement mathematique, 36, 267 (1990).
\item[[7]] F.D.M. Haldane, Phys. Rev. Lett.  60, 635(1988);S. Shastry, Phys. Rev. Lett. 60, 639 (1988);Haldane F.D.M., Physics of the ideal semion gas: Spinons and quantum symmetries of the integrable Haldane-Shastry spin chain,
Proceedings of the 16th Tauiguchi Symposium on Condensed Matter Physics, Kashikojima, Japan, ed. by O. OKiji and N. Kawakami, Springer, Berlin, 1994. 
\item[[8]] M.L. Ge, K. Xue,and Y.M. Cho, Phys. Lett. A249, 258 (1998);C.M. Bai, M.L. Ge and K. Xue, Directional Transitions in spin systems and representations of $Y(sl(2))$, Nankai preprint, APCTP-98-026.
\item[[9]] C.M. Bai, M.L. Ge and K. Xue, J. Stat. Phys. 102, 545 (2001).
\item[[10]] A.J. Leggett, Rev. Mod. Phys. 47, 331 (1975); G.E. Volovik, Exotic properties of superfluid $^3$He, World Scientific, Singapore, 1992.
\end{description}

\end{document}